ORIGINAL ARTICLE

# COVID-19 and income profile: How communities in the United States responded to mobility restrictions in the pandemic's early stages

**Qianqian Sun** | **Weiyi Zhou** | **Aliakbar Kabiri** | **Aref Darzi** | **Songhua Hu** | **Hannah Younes** | **Lei Zhang**

Maryland Transportation Institute (MTI), Department of Civil and Environmental Engineering, University of Maryland, 8228 Paint Branch Dr, College Park, MD, 20742, USA

**Correspondence**
Qianqian Sun, Maryland Transportation Institute (MTI), Department of Civil and Environmental Engineering, 8228 Paint Branch Dr, University of Maryland, College Park, MD 20742, USA.
Email: qsun12@umd.edu

## Abstract

Mobility interventions in communities play a critical role in containing a pandemic at an early stage. The real-world practice of social distancing can enlighten policymakers and help them implement more efficient and effective control measures. A lack of such research using real-world observations initiates this article. We analyzed the social distancing performance of 66,149 census tracts from 3,142 counties in the United States with a specific focus on income profile. Six daily mobility metrics, including a social distancing index, stay-at-home percentage, miles traveled per person, trip rate, work trip rate, and non-work trip rate, were produced for each census tract using the location data from over 100 million anonymous devices on a monthly basis. Each mobility metric was further tabulated by three perspectives of social distancing performance: "best performance," "effort," and "consistency." We found that for all 18 indicators, high-income communities demonstrated better social distancing performance. Such disparities between communities of different income levels are presented in detail in this article. The comparisons across scenarios also raise other concerns for low-income communities, such as employment status, working conditions, and accessibility to basic needs. This article lays out a series of facts extracted from real-world data and offers compelling perspectives for future discussions.

**KEYWORDS**
causal impact analysis of income, community differences, COVID-19 pandemic, Mobile location data, social distancing behavior

**JEL CLASSIFICATION**
I180, L910, L980, R230, R400

## 1 | INTRODUCTION

On March 11, 2020, the World Health Organization declared COVID-19, a novel coronavirus, a global pandemic. Since this pandemic was the largest public health emergency in the last century, it had an unparalleled impact on society in terms of public health, economy, education, environment, and psychology (Almeida et al., 2021; Baker et al., 2020; Miyah et al., 2022). Even though it has been two years since this global crisis first emerged, research about it is still valuable. It has pertinent lessons that can be applied in the future to assist in combatting the next potential comparable pandemic. In the early stages of the pandemic, when there are no vaccinations or effective therapies, mobility intervention, a non-pharmaceutical strategy, is crucial in slowing the virus's rate of transmission. Measuring the effectiveness of mobility interventions is important for improving policies.

Our article investigates how communities in the United States responded to mobility restrictions in the early stages of the pandemic. To recap, the US government declared a national emergency on March 13, 2020, and that same day, a proclamation prohibiting noncitizens from entering the country went into force. Later, the Coronavirus Guidelines for America were released on March 16, 2020, in which mobility restrictions were strongly recommended (Brzezinski et al., 2020; Sadique et al., 2007). The first stay-at-home order in the U.S. began in California on March 19 and swiftly spread across the country. All but eight states had stay-at-home orders in place by mid-April. However, it was discovered that despite two months having elapsed since the national emergency declaration, COVID-19 cases had continued to rise alarmingly quickly. According to our study, there is much room for improvement in early mobility intervention. For greater success, these initiatives should not be implemented as a standalone policy but rather as a collaborated effort among many societal sectors.

According to past studies, a number of social and economic factors are positively correlated with the transmission of COVID-19 (Qiu et al., 2020; Sirkeci & Yucesahin, 2020). This article concentrates on the income profile of local communities and provides empirical evidence or reference for related research and decision-makers. Different from the general meaning of reducing physical contact between people, "social distancing" in this article refers to a decrease in mobility.

Our article examines how income influences the communities' performance of social distancing. The bottleneck for achieving social distancing is discovered to be low-income neighborhoods, and the underlying causes are examined. Additionally, policymakers, when making decisions, should pay attention to the problems and difficulties that low-income communities encountered. They should consider the low capacity of low-income people to cope with pandemics due to the threat of unemployment, exposure in the working environment, low accessibility to daily supplies, and limited access to healthcare. This is supported by our findings in terms of daily mobility patterns.

Social inequality is a perennial problem in the sociology field. The needs assessment of underserved and disadvantaged communities and making plans to optimize resource allocation are encouraged by our article. This article also provides insights into current policies. For instance, telecommuting is a beneficial substitution for working on site. It should be applied to more industries so that more low-income people can also work from home. The pandemic also changed the education system. Parents play a role in promoting children's engagement in online courses. While the relatively higher daily travel needs, both work and non-work related, of low-income communities as revealed in this article should be considered by educators. For underprivileged populations, customized teaching plans or special education benefits may be offered. Additionally, this article highlights the significance of mobile location data in capturing the public's daily travel patterns, which is also useful for tracking and predicting the pandemic trends. Mobile location data is now more widely available and can be utilized by organizations in various fields.

In this article, communities, specifically at census tract level, are carefully examined in terms of their social distancing performance from daily mobility perspective. Social distancing is measured by daily mobility metrics that quantify personal travel statistics. The less frequent or the shorter distance a person travels, the less likely they are to have contact with others. Six daily mobility metrics are therefore compared between income groups, including the social distancing index (SDI), staying-home percentage, miles traveled per person, trip rate, work trip rate, and non-work trip rate. Each metric is further tabulated by "best performance", "effort", and "consistency" (definitions are in the methodology section). A total of 18 indicators are reviewed in order to evaluate the impact of income on a community's social distancing performance, which provide answers to the following questions: what are the differences in mobility metrics between high- and low-income communities, do communities with different income levels behave differently in terms of social distancing in a statistically relevant way, and how does income influence the 18 indicators. We find a temporal difference between the two income groups in each mobility metric, and hypothesis tests show a substantial difference between the two groups in the most of the 18 indicators. Furthermore, propensity score analysis is used to examine how income influences the 18 indicators.

Mobile location data, an emerging data source for analyzing mobility features, has been extensively utilized to investigate public mobility throughout the pandemic (Boyd et al., 2006; Ghader et al., 2020; Lee et al., 2020; Pan et al., 2020; Xiong et al., 2020; Zhang et al., 2020). With access to high-frequency nationwide real-world mobile location data integrated from over 100 million anonymous devices on a monthly basis, this article's authors were able to develop an analysis of 66,149 census tracts from 3,142 counties in the U.S. We explored how these census tracts (that is, communities) performed regarding social distancing from a mobility aspect. The temporal mobility patterns of different income groups are first analyzed through repeated measures analysis of variance (ANOVA) and post hoc analysis, which statistically indicate the considerable differences for pairwise comparisons between days (Samuel & Jessica, 2001; Souraya & Lynn, 1993; Sun et al., 2020). Based on this, we hypothesized high-income communities outperformed low-income communities in social distancing performances, which is corroborated by the results of Welch's *t*-test and propensity score modelling (Caliendo & Kopeinig, 2008; Guo & Fraser, 2014; Shipman et al., 2017). Our study outlines the detailed differences in social distancing between high- and low-income communities. Some intriguing findings are revealed, which not only provide policymakers with reasons to take better precautions during pandemics but also shed important insights into human behavior research in social science and mobility analysis in the transportation area.

## 2 | METHODOLOGY

### 2.1 | Deriving the indicators of "social distancing" performance

Mobile location data have been actively employed in mobility analysis for their benefits of real-world representation, high location accuracy, high spatial and temporal resolution, large spatial and temporal coverage, and real-time availability. To generate the daily mobility metrics of communities in the U.S., a multi-sourced passively collected mobile location dataset is utilized, from which the movements of over 100 million devices across the nation in a month are observed. This dataset provides the location (latitude and longitude) and coordinated universal time (UTC) timestamp for each anonymized device identifier. Since the dataset is passively collected, it is not directly usable for mobility analysis. Hence, the authors developed a set of algorithms for deriving the daily mobility metrics (Pan et al., 2020;

Zhang, Darzi, Ghader, et al., 2021). The same algorithms are applied in this article to compute the mobility metrics (SDI, staying-home percentage, miles traveled per person, trip rate, work trip rate, and non-work trip rate) specifically at census tract level. Each metric is tabulated by three newly defined perspectives ("best performance", "effort", and "consistency") so that 18 indicators, as a reflection of social distancing performance, are analyzed. Their definitions are summarized in Table 1. The listed periods in Table 1 are defined based on our previous findings (Lee et al., 2020; Sun et al., 2020; Zhang, Darzi, Ghader, et al., 2021), which are also exemplified in this article's results section. Weekends and federal holidays are excluded throughout this article. Taking SDI as an example, Figure 1 displays the SDI changes across the four periods.

Three steps are involved for producing the 18 indicators. The first step is applying a pipeline designed by the authors to produce a national trip profile. The second step is generating the six daily mobility metrics from the national trip profile. The third step is tabulating each metric by best performance, effort, and consistency.

Since a trip is the basic travel analysis unit of mobility behavior, building the national trip profile is necessary, which contains device- and trip-level attributes, including home location of devices, trip origin and destination information (UTC timestamp, latitude and longitude), trip distance, trip purpose, and trip weight. All the algorithms involved in the pipeline are peer-reviewed and have been published in the other articles by the authors (Zhang, Darzi, Ghader, et al., 2021) and reports (Zhang et al., 2020; Zhang, Darzi, Pan, et al., 2021). The pipeline's major steps are as follows. First, the multi-sourced mobile location data go through a data-cleaning process (data oscillation identification and removal, quality control criteria) and a data-fusion process (deduplication). Second, an algorithm is implemented to construct the home location profile for the mobile devices. Third, a recursive algorithm is applied to identify trips for each mobile device. Fourth, trip attributes, including trip purpose (work and non-work) and trip distance are imputed. Lastly, a multilevel weighting algorithm is applied to address the sample biases and to expand the sample trip profile to make it representative at the population level. Details on each algorithm can be found in the aforementioned articles and reports.

**TABLE 1** Summary of definitions

| | Term | Definition |
|---|---|---|
| Mobility metrics | Social distancing index (SDI) | A 0–100 index that measures the extent of social distancing behavior, with 0 meaning the lowest extent and 100 meaning the highest, whose computation involves other five mobility metrics and out-of-county trips. |
| | Staying-home percentage | Percentage of residents staying at home without conducting any trip longer than one mile from home. |
| | Miles traveled per person | Average person miles traveled (PMT) on whatever travel modes (walking, biking, driving, transit, air, etc.). |
| | Trip rate | Average daily number of trips per person. |
| | Work trip rate | Average daily number of work trips (ending at the workplace) per person. |
| | Non-work trip rate | Average daily number of non-work trips (ending at places other than the workplace, e.g., supermarkets or gas stations,) per person. |
| Periods | Pre-pandemic | The two weeks before March 13, 2020 (i.e., March 2–March 12, 2020). |
| | Behavior-change | The period that people quickly changed mobility behavior in response to stay-at-home orders (i.e., March 13–March 27, 2020). |
| | Inertia | The period showing that mobility metrics gradually stabilized (i.e., March 30–April 10, 2020, excluding the weekend March 28–March 29, 2020). |
| | Fatigue | The week of April 13, 2020, with each metric showing a slight rebounding trend despite ongoing stay-at-home orders (i.e., April 13–April 17, 2020). A week after the fatigue period, states successively began partially reopening local businesses. |
| Perspectives | Best performance | The most expected value of a mobility metric during the behavior-change, inertia, and fatigue period. For example, the maximum SDI value (the higher the value, the more socially distanced) and the minimum value of trip rate are referred to as best performance. |
| | Effort | The percentage change (positive or negative) in each metric by comparing the best performance with the baseline—the weekday average in the pre-pandemic period. |
| | Consistency | The standard deviation of each mobility metric in the inertia and fatigue periods. |

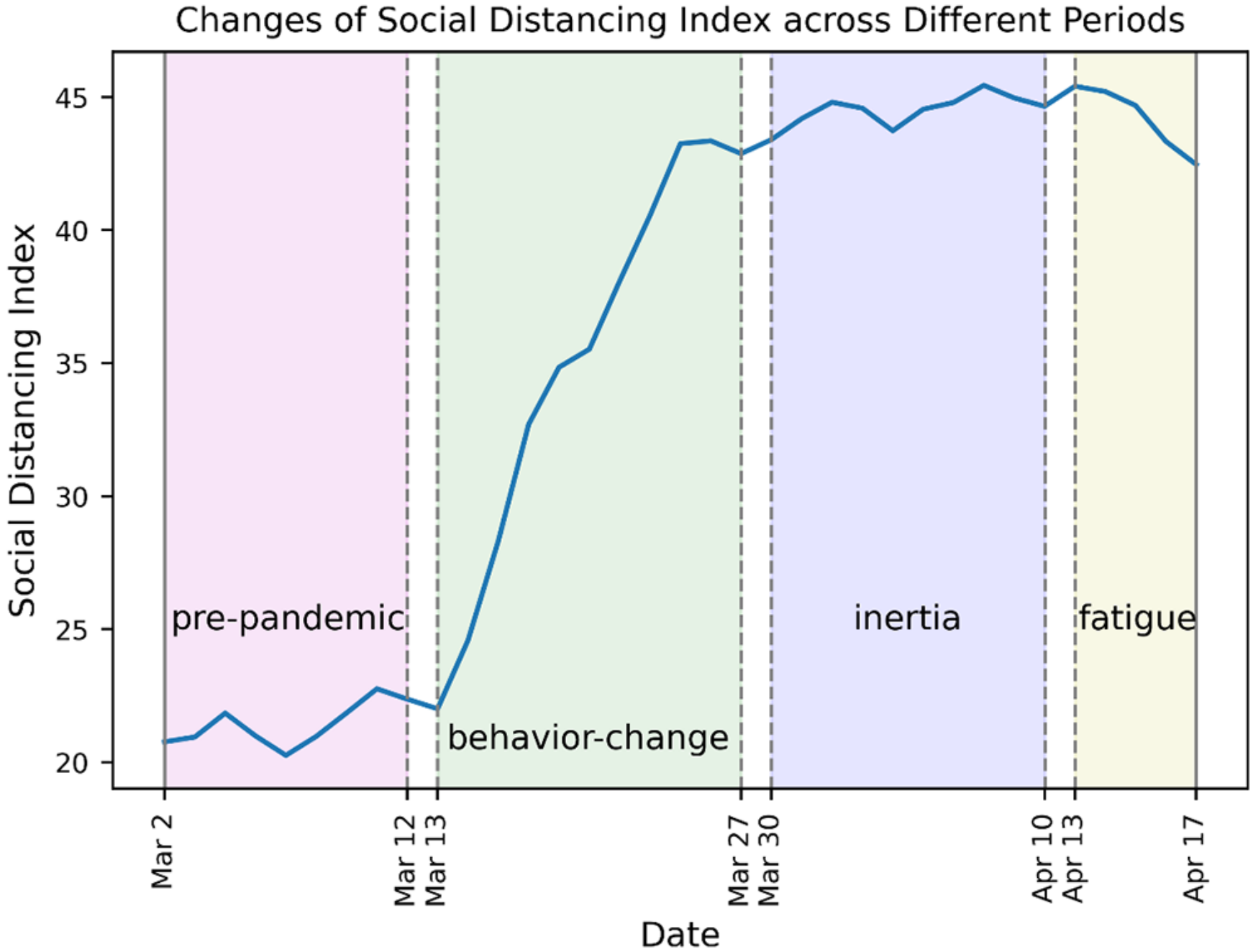

**FIGURE 1** SDI changes across different periods

After producing the national trip profile, the six mobility metrics are imputed at census tract level for this article's analysis. The imputed metrics at higher levels (national, state, and county) are published on a platform (Maryland Transportation Institute, 2020).

## 2.2 | Defining high- and low-income communities

In line with the US Internal Revenue Code Section 45D(e) (1), we classify any census tract with at least a 20% poverty rate or median household income of no more than 80% of the metropolitan area or statewide median income as a low-income community. The income data are obtained from the American Community Survey (ACS) five-year 2015–2019 estimate to classify the census tracts into high- and low-income groups.

## 2.3 | Preliminary analysis and hypothesis testing

In the preliminary analysis, income groups' temporal patterns of mobility metrics were discussed and further investigated through repeated measures one-way ANOVA (RM-ANOVA) and post hoc analysis. RM-ANOVA tests whether the pairwise differences between high and low-income communities on each day are statistically significant. Post hoc analysis additionally demonstrates which time points significantly differ from others. Based on the findings from the preliminary analysis, we propose the hypothesis that high-income communities outperformed low-income communities in three aspects: (1) high-income communities showed better best performance, (2) high-income communities made more effort in social distancing, and (3) high-income communities displayed a higher consistency of practicing social distancing. Based on these three aspects and the six mobility metrics, we employ 18 indicators to compare the high- and low- income communities. One-sided Welch *t*-tests are applied to the 18 indicators to test one common hypothesis that high-income communities outperformed low-income communities. When comparing the best performance and efforts, the alternative hypothesis used for staying-home percentage and SDI is that high-income communities' mean is greater than low-income communities' mean, while for the other four mobility metrics the reverse is true. When comparing the consistency, the alternative hypothesis used for all mobility metrics is that high-income communities' mean is smaller than low-income communities.

## 2.4 | Quantifying the impact of income on social distancing performance

We additionally conduct an impact analysis of income on mobility pattern. Considering the systematic biases between the two income groups probably caused by covariates (Austin, 2011; Dehejia & Wahba, 2002), we apply a propensity score matching (PSM) method. PSM has been popularly applied to estimate the treatment effect in observational studies (Caliendo & Kopeinig, 2008; Guo & Fraser, 2014; Shipman et al., 2017). When implementing PSM, high income is the treatment. Four covariates regarding age, sex, race, and education, which are expected to influence the income level (treatment) and social distancing performance (outcome) based on studies (Engle et al., 2020; Fang et al., 2020; Xiong et al., 2020), are set as control variables. They are obtained from the ACS five-year 2015–2019 estimate. The PSM process consists of three steps: producing the propensity score of being high income for all communities, matching the high-income and low-income communities, and imputing the average treatment effect of being high income. This process is repeated for the 18 indicators to measure the income's impact on communities' social distancing performance.

First, a binomial logistic model—Equations (1) and (2)—is built to estimate the propensity score (i.e., the probability of being high income conditioning the four covariates) of a community.

$$\frac{e^{\beta_0+\beta_1 X_1+\beta_2 X_2+\beta_3 X_3+\beta_4 X_4}}{1+e^{\beta_0+\beta_1 X_1+\beta_2 X_2+\beta_3 X_3+\beta_4 X_4}}; \tag{1}$$

$$\text{logit}(P) = \ln\left(\frac{P}{1-P}\right), \tag{2}$$

where $X_1$, $X_2$, $X_3$, $X_4$ are the percentage of people aged 65 and over, the percentage of males, the percentage of Black or African American people, and percentage of people with a high school education or less respectively; P is the propensity score.

Based on all the communities' estimated propensity score, the nearest neighbor method is applied to match high-income and low-income communities. Taking a high-income community, the community from the low-income group with the closest propensity score is selected as the counterpart. This process forms a set of high- and low-income community pairs with each pair having the two counterfactual values.

Finally, a less biased treatment impact is measured for each community by the difference between the two counterfactual values (Ashenfelter, 1978). Then the average treatment effect (ATE) across all communities and the average treatment effect on the treated (ATET) only across high-income communities are used to measure the expected impact of being high income.

## 3 | RESULTS

This section first summarizes the preliminary findings that present the temporal differences in the mobility patterns between income groups. Then, it presents the hypothesis-testing results indicating that high-income communities showed better compliance with government guidelines than low-income communities in most cases. Lastly, it shows the impact analysis results of income on the mobility pattern, which indicate that being high income promotes better social distancing performance at the community level.

### 3.1 | Preliminary results on the temporal differences between income groups

The temporal patterns of the six mobility metrics by low- and high-income groups are summarized. Figure 2 exemplifies the staying-home percentage by census tract on March 13, 2020, and April 13, 2020. The April 13 map shows a much higher staying-home percentage across the nation than the March 13 map.

Figure 3 below shows the temporal changes of the mean value of mobility metrics by high- and low-income groups from March 2, 2020, to April 17, 2020. It is obvious that both income groups experienced the same two critical time points—March 13, 2020, and April 13, 2020—as found in our previous studies at the state and county levels (Lee et al., 2020; Sun et al., 2020; Zhang, Darzi, Ghader, et al., 2021).

When comparing the two income groups, notable differences in all the mobility metrics throughout the entire study period (March 2–April 17, 2020) are observed. Interestingly, the relative positions of the two groups swapped regarding SDI, staying-home percentage, and miles traveled per person (Figure 3). At beginning, high-income communities seemed to face more challenges in following social-distancing orders with a lower SDI, a lower staying-home percentage, and a higher miles traveled per person. However, they quickly responded and surpassed low-income communities soon after March 13. As for the trip rate and non-work trip rate (Figure 3), during the pre-pandemic period, there was already a difference between the two groups, and this difference kept shrinking. However, after March 13, the difference suddenly stopped shrinking; instead, it began growing and later stabilized. A greater difference than existed pre-pandemic between the groups developed. As for work trip rate, the disparity after March 13 also became much more obvious. Overall, high-income communities made more efforts to socially distance than low-income communities, meaning high-income communities had higher absolute value of percentage changes in all six metrics. High-income communities were at a lower risk of spreading COVID-19 indicated by all six mobility metrics.

Moreover, with RM-ANOVA, we investigate the temporal variation of three mobility metrics (work trip rate, staying-home percentage, and miles traveled per person) statistically. These three metrics are selected because they represent three different temporal patterns, as shown in Figure 3. Work trip rate, trip rate, and non-work trip rate share the same pattern whereby the values for the two income groups decrease overall without intersection. Moreover, the staying-home percentage and SDI share the same pattern whereby the values for the two income groups increase overall and intersect at some point. Meanwhile, for miles traveled per person, the trend for both income groups overall decreases and intersects at some point. We sorted the census tracts based on their median income and chose the top 200 and the bottom 200 census tracts for more detailed comparison. The results of the three metrics for both groups are presented in the significance plots (Figure 4), which reveal the following findings. First, significant change of means at a 99.9% confidence interval, denoted by the dark green cells, are observed earlier in high-income communities regarding all three metrics. This indicates that high-income communities responded earlier to the mobility interventions. For example, high-income communities responded 10 days earlier in terms of work trips per person and miles traveled per person than low-income communities (Figures 4a, c, d, and f). Second, high-income communities finished changing their behavior more quickly with regards to the staying-home percentage and miles traveled per person than low-income communities. This is supported by the smaller and clearer pink square observed in the middle of the significance plots of the high-income group (Figures 4b, c, e, and f). Lastly, high-income communities' social distancing performance is more consistent, long-lasting, and stable in all three metrics. After the behavior-change period, high-income people kept socially distancing, while low-income communities showed more fluctuations.

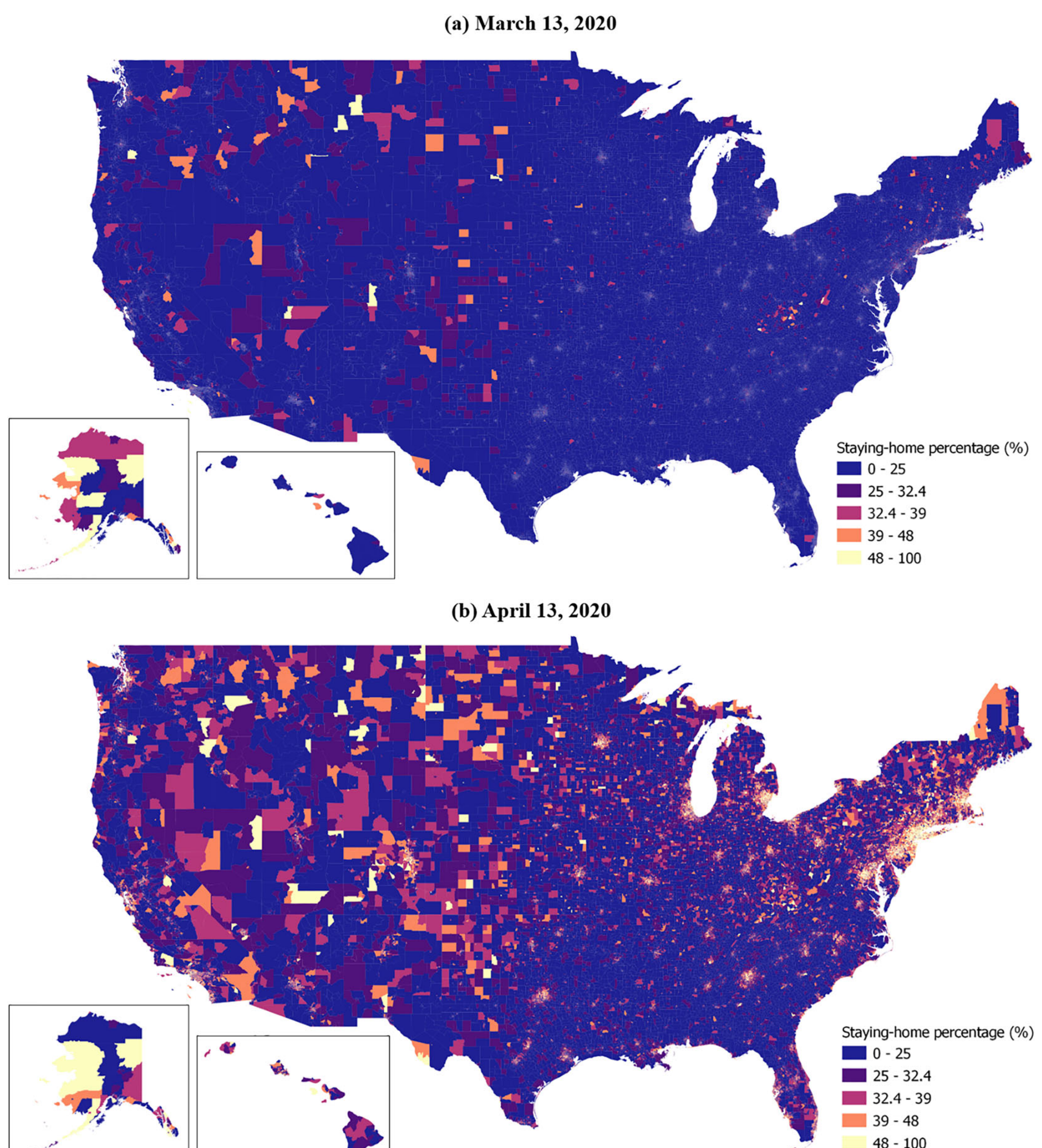

**FIGURE 2** Staying-home percentage by census tracts on March 13, 2020, and April 13, 2020

## 3.2 | Hypothesis-testing results

The previous section elucidates the differences between income groups at the average level. In this section, the difference is further detailed through distribution plots by income groups (Figure 5). Generally, the distributions of both groups are close to normal distribution, and high-income communities tend to perform better. As for best performance, the SDI and staying-home percentage means for high-income communities is larger than that of low-income communities, while the mean of the other metrics for high-income communities is smaller. As for effort, on average high-income communities made more effort since the mean absolute value of the percentage change in all six metrics for high-income communities is larger than that for low-income communities. Regarding consistency, high-income communities have a smaller standard deviation, indicating that they were more stable when socially distancing in the inertia and fatigue periods.

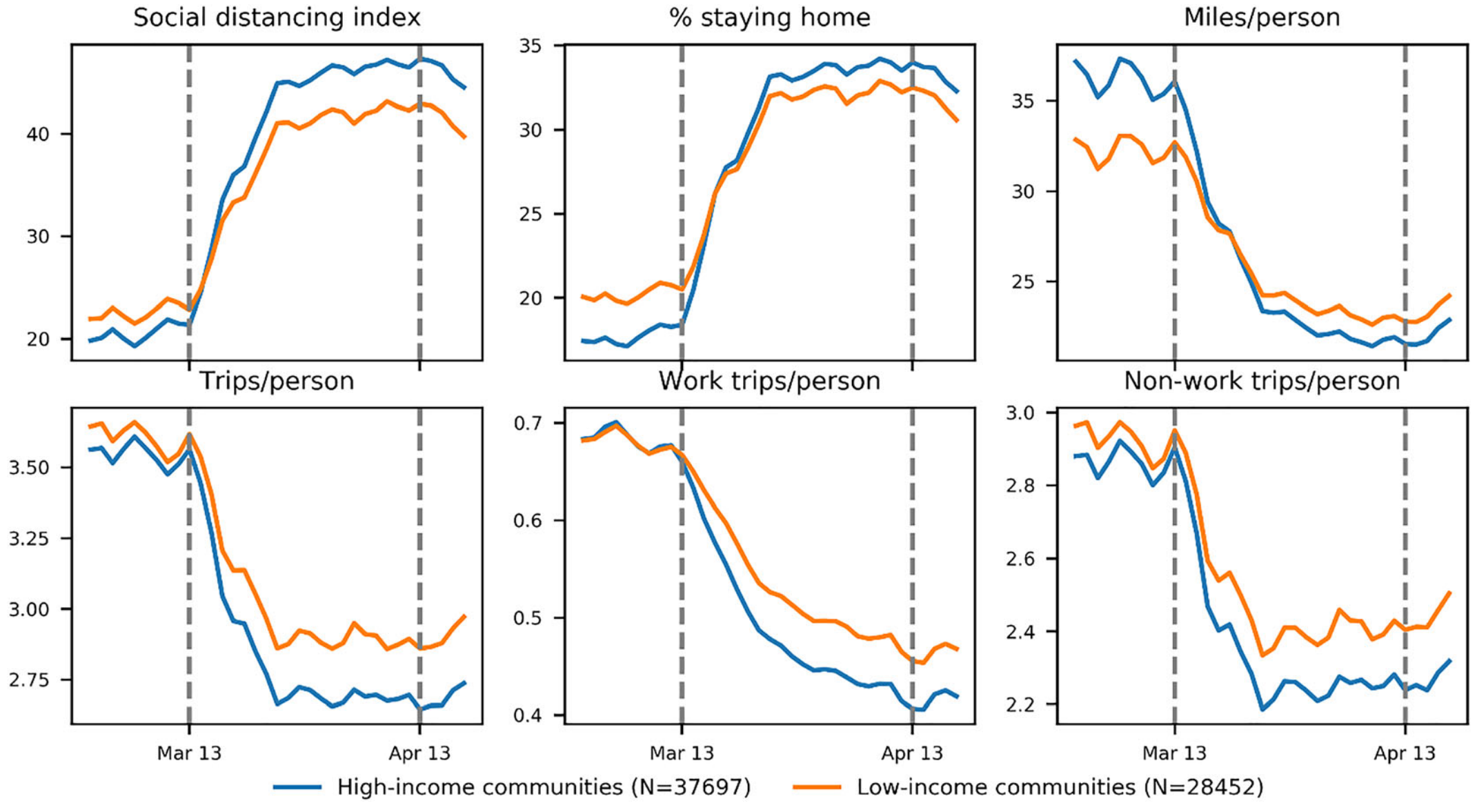

**FIGURE 3** Temporal patterns of six mobility metrics from March 2–April 17, 2020, by income groups

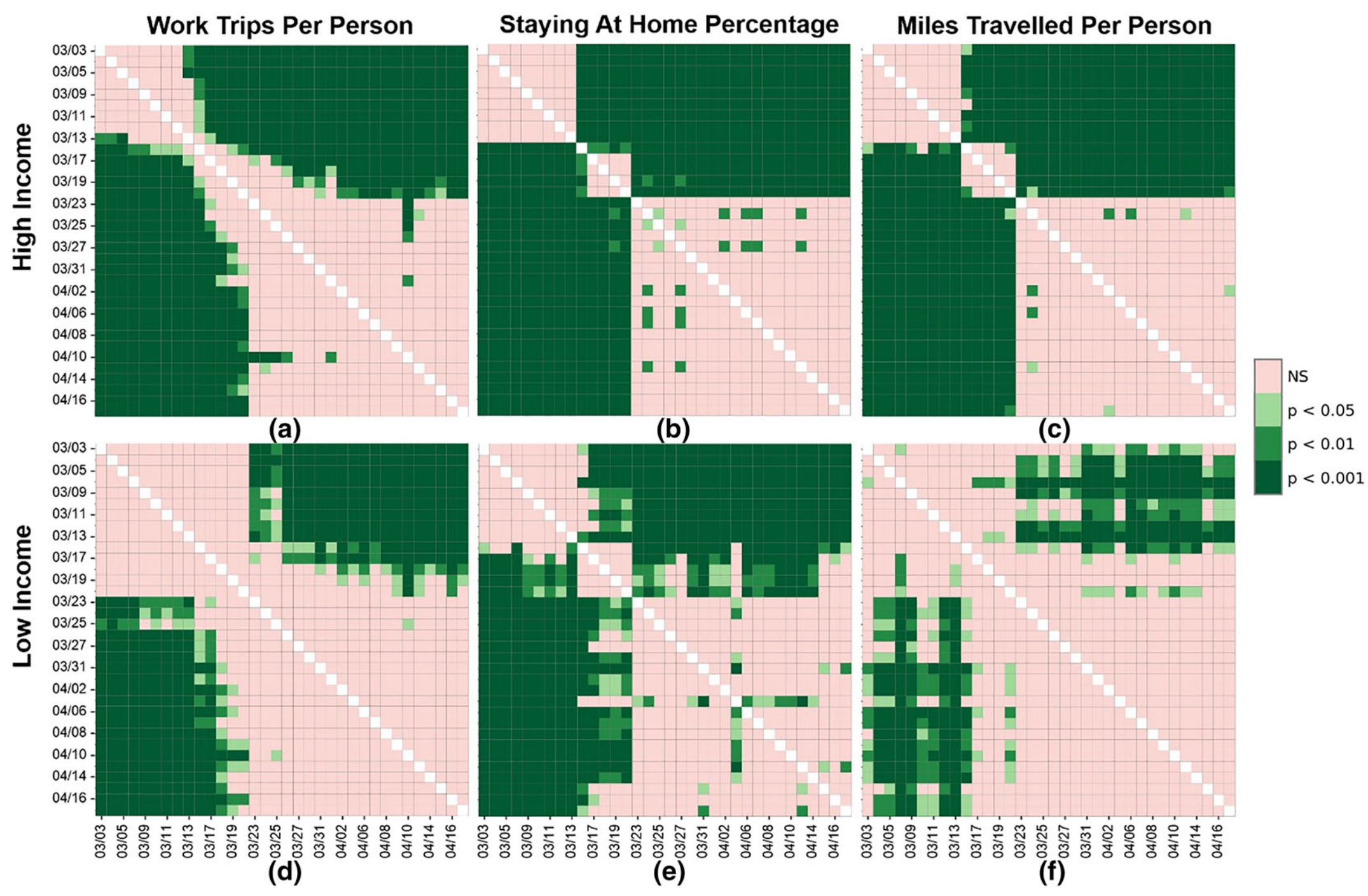

**FIGURE 4** The disparities in course progression over time of three mobility metrics

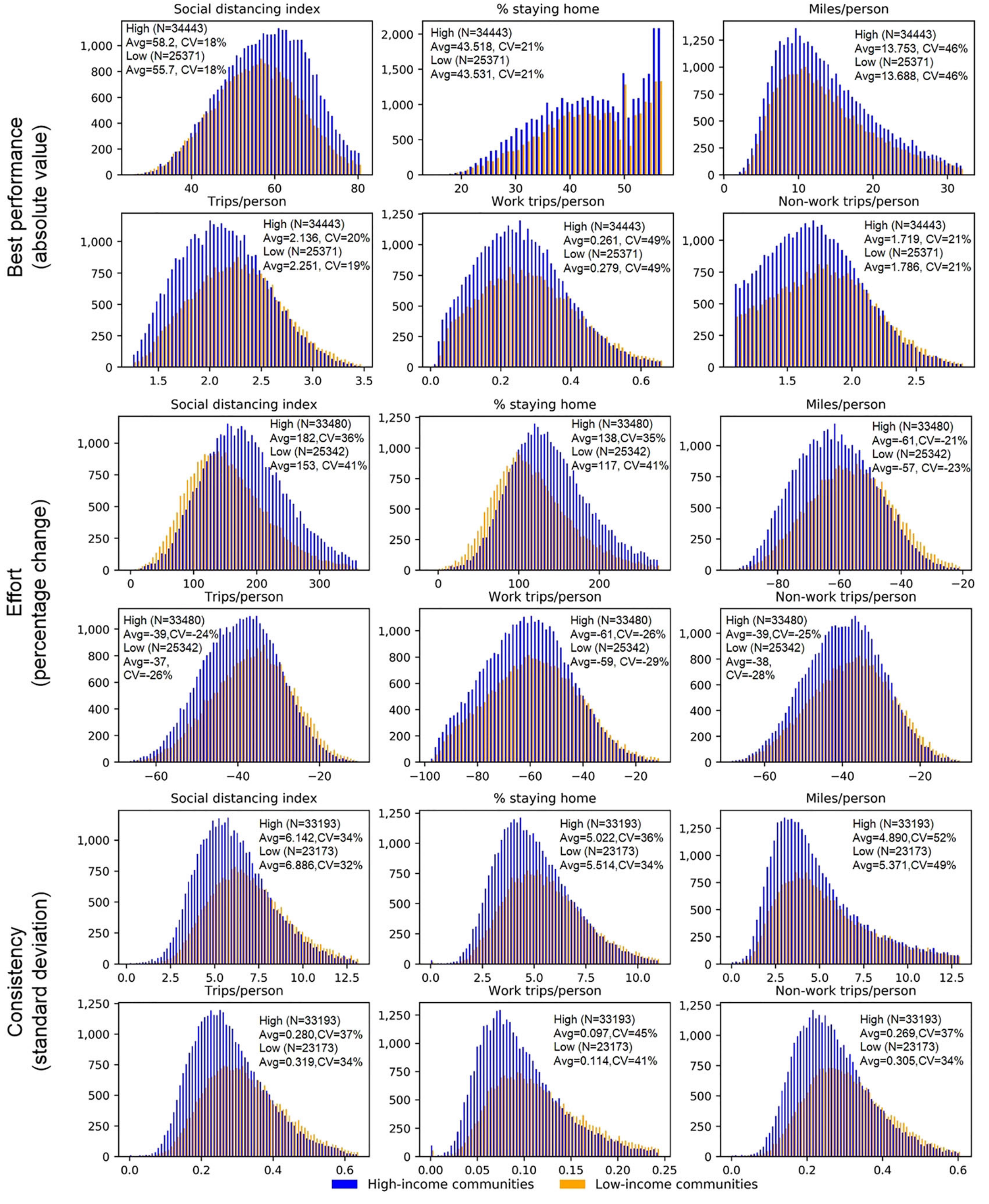

**FIGURE 5** Distribution of the social distancing performance by income groups

The results of one-sided Welch *t*-tests for all scenarios are summarized in Table 2, showing the alternative hypotheses of 16 scenarios are supported at the 99% confidence level. As for best performance, the SDI mean is

**TABLE 2** One-sided Welch $t$-test results for 18 scenarios considering three types of indicators and six mobility metrics

| Group | Mobility metrics | $\mu_1$ | $\mu_2$ | $H_a$ | Df | $t$ (sig.) | 95% CI | |
|---|---|---|---|---|---|---|---|---|
| Best performance (extremum) $N_1=34{,}443$, $N_2=25{,}371$ | Social distancing index | 58.1 | 55.7 | $\mu_1>\mu_2$ | 55,010 | $28.6^{***}$ | 2.3 | Inf |
| | % staying home | 43.5 | 43.5 | $\mu_1>\mu_2$ | 55,476 | −0.17 | −0.14 | Inf |
| | Miles traveled/ person | 13.8 | 13.7 | $\mu_1<\mu_2$ | 54,710 | −1.26 | −Inf | 0.15 |
| | Trips/person | 2.14 | 2.25 | $\mu_1<\mu_2$ | 54,054 | $-32.9^{***}$ | −Inf | −0.11 |
| | Work trips/ person | 0.26 | 0.28 | $\mu_1<\mu_2$ | 52,631 | $-16.3^{***}$ | −Inf | −0.02 |
| | Non-work trips/ person | 1.72 | 1.79 | $\mu_1<\mu_2$ | 52,809 | $-21.9^{***}$ | −Inf | −0.06 |
| Effort (percentage change) $N_1=33{,}480$, $N_2=25{,}342$ | Social distancing index | 182 | 153 | $\mu_1>\mu_2$ | 55,624 | $55.4^{***}$ | 28.5 | Inf |
| | % staying home | 138 | 117 | $\mu_1>\mu_2$ | 54,550 | $53.9^{***}$ | 21.0 | Inf |
| | Miles traveled/ person | −61 | −57 | $\mu_1<\mu_2$ | 53,731 | $-40.9^{***}$ | −Inf | −4.3 |
| | Trips/person | −39 | −37 | $\mu_1<\mu_2$ | 53,627 | $-29.6^{***}$ | −Inf | −2.2 |
| | Work trips/ person | −61 | −59 | $\mu_1<\mu_2$ | 53,053 | $-15.8^{***}$ | −Inf | −2.0 |
| | Non-work trips/ person | −39 | −38 | $\mu_1<\mu_2$ | 52,826 | $-16.4^{***}$ | −Inf | −1.3 |
| Consistency (standard deviation) $N_1=33{,}193$, $N_2=23{,}173$ | Social distancing index | 6.14 | 6.89 | $\mu_1<\mu_2$ | 48,619 | $-40.6^{***}$ | −Inf | −0.71 |
| | % staying home | 5.0 | 5.5 | $\mu_1<\mu_2$ | 48,549 | $-30.7^{***}$ | −Inf | −0.47 |
| | Miles traveled/ person | 4.89 | 5.37 | $\mu_1<\mu_2$ | 48,661 | $-21.6^{***}$ | −Inf | −0.44 |
| | Trips/person | 0.280 | 0.319 | $\mu_1<\mu_2$ | 48,330 | $-42.8^{***}$ | −Inf | −0.04 |
| | Work trips/ person | 0.097 | 0.114 | $\mu_1<\mu_2$ | 47,472 | $-45.1^{***}$ | −Inf | −0.02 |
| | Non-work trips/ person | 0.269 | 0.305 | $\mu_1<\mu_2$ | 48,359 | $-41.0^{***}$ | −Inf | −0.03 |

*Note*: 1. High-income group mean: $\mu_1$, low-income group mean: $\mu_2$, 2. significance level: 0.0001 - '***', 0.1 - '', 3. $H_a$: alternative hypothesis.

statistically significantly higher in high-income communities than in low-income areas, while the average trip rate, work trip rate, and non-work trip rate are all statistically significantly lower in high-income communities than in low-income communities. As for effort, a significant difference in mean is seen for the staying-home percentage and SDI. Although the two income groups show very close percentage changes in the other four metrics, high-income communities still significantly outperform low-income communities. The largest difference is observed in the miles traveled per person (<= −4.3%), and the difference in work trip rate (<= −2.0%) is larger than in non-work trip rate (<= −1.3%). Overall, high-income communities made more efforts to practice social distancing. Lastly, we compared the two groups regarding their social distancing consistency during the inertia and fatigue periods. Result shows that the mean value for consistency in high-income communities is statistically lower than for low-income communities for all six metrics. After the behavior-change period, high-income communities more constantly and consistently socially distanced. It is noted that the best performance of both the staying-home percentage and miles traveled per person are not significantly different between income groups. It turns out that high-income and low-income groups presented the maximum staying-home percentage and the minimum miles traveled per person on the same day, April 13, 2020 (the start of the fatigue period), indicating the two income groups finally reduced their travels to a similar extent from these two aspects and if they were not growing fatigued, high-income communities probably could have performed better.

## 3.3 | Influence of income on social distancing performance

### 3.3.1 | Propensity score and matching

A binomial logistic regression analysis is conducted using Equations (1) and (2) to produce the propensity score of being high income for all communities with four covariates: percentage of people aged 65 and over, percentage of males, percentage of Black or African American people, and percentage of people with a high school education or less. Table 3 summarizes the logistic regression results, which show that the four variables effectively influence the probability of being high income. By adding the four independent variables to the intercept, goodness of fit is improved. McFadden's $R^2$ is between 0.2 and 0.4 meaning an excellent model fit (McFadden, 1979).

Since there are more high-income communities, nearest neighbor matching with a replacement is applied. The distribution of propensity score in the matched treated group and the matched control group in shown in the Figure 6. The absolute standardized mean difference comparison of before and after matching shows that the balance on the covariates is overall much improved, and all four covariates' absolute standardized mean difference is within the threshold of 0.1 after matching (Figure 7).

### 3.3.2 | Measuring income's impact on social distancing performance

Table 4 summarizes the ATE and ATET for the 18 scenarios. It shows that high income has a statistically significant causal impact on social distancing performance at a 99.9% confidence level in all 18 scenarios. Also, the causal

**TABLE 3** Binomial logistic regression results

| | Estimate | Std. error | *z* value | $\Pr(>\lvert z\rvert)$ | Odds ratio | 95% C.I. for odds ratio | VIF |
|---|---|---|---|---|---|---|---|
| Intercept | 2.886 | 0.157 | 18.387 | <2e-16 *** | / | / | / |
| Percentage of people aged 65 and over | 0.064 | 0.002 | 34.763 | <2e-16 *** | 1.066 | [1.062, 1.070] | 1.092 |
| Percentage of male people | 0.011 | 0.003 | 3.601 | 0.0003 *** | 1.011 | [1.005, 1.017] | 1.084 |
| Percentage of Black or African American people | −0.033 | 0.001 | −50.502 | <2e-16 *** | 0.968 | [0.966, 0.969] | 1.069 |
| Percentage of people without a higher education degree | −0.097 | 0.001 | −106.546 | <2e-16 *** | 0.908 | [0.906, 0.909] | 1.021 |
| Null deviance | 81,539 on 59,813 degrees of freedom | | | | | | |
| Residual deviance | 53,271 on 59,809 degrees of freedom | | | | | | |
| McFadden's $R^2$ | 0.35 | | | | | | |

*Note*: 1. significance level: 0 '***' 0.001 '**' 0.01 '*' 0.05 '.' 0.1 ' '; 2. N=59,814 cases, AIC=53281.

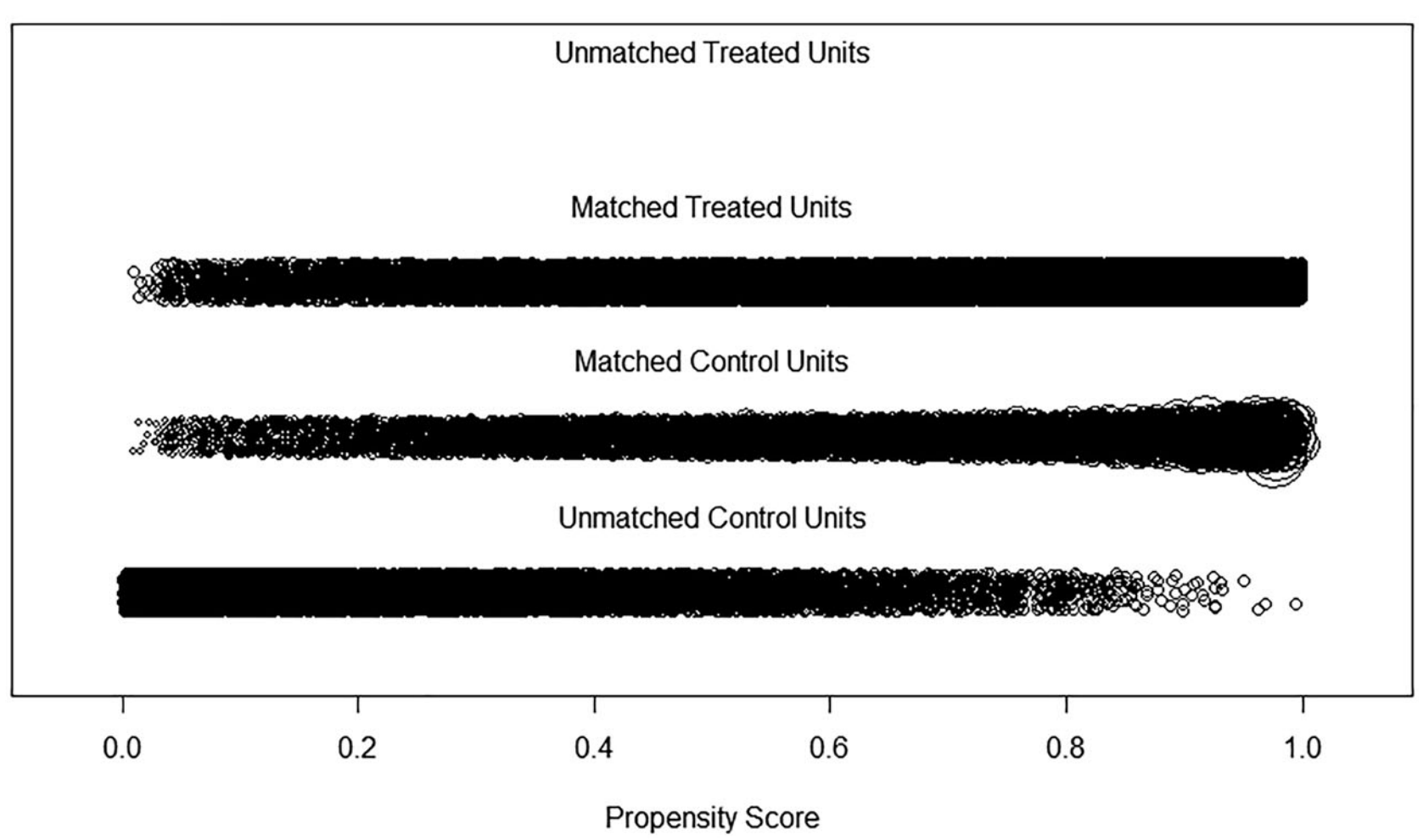

**FIGURE 6** Distribution of propensity scores by groups

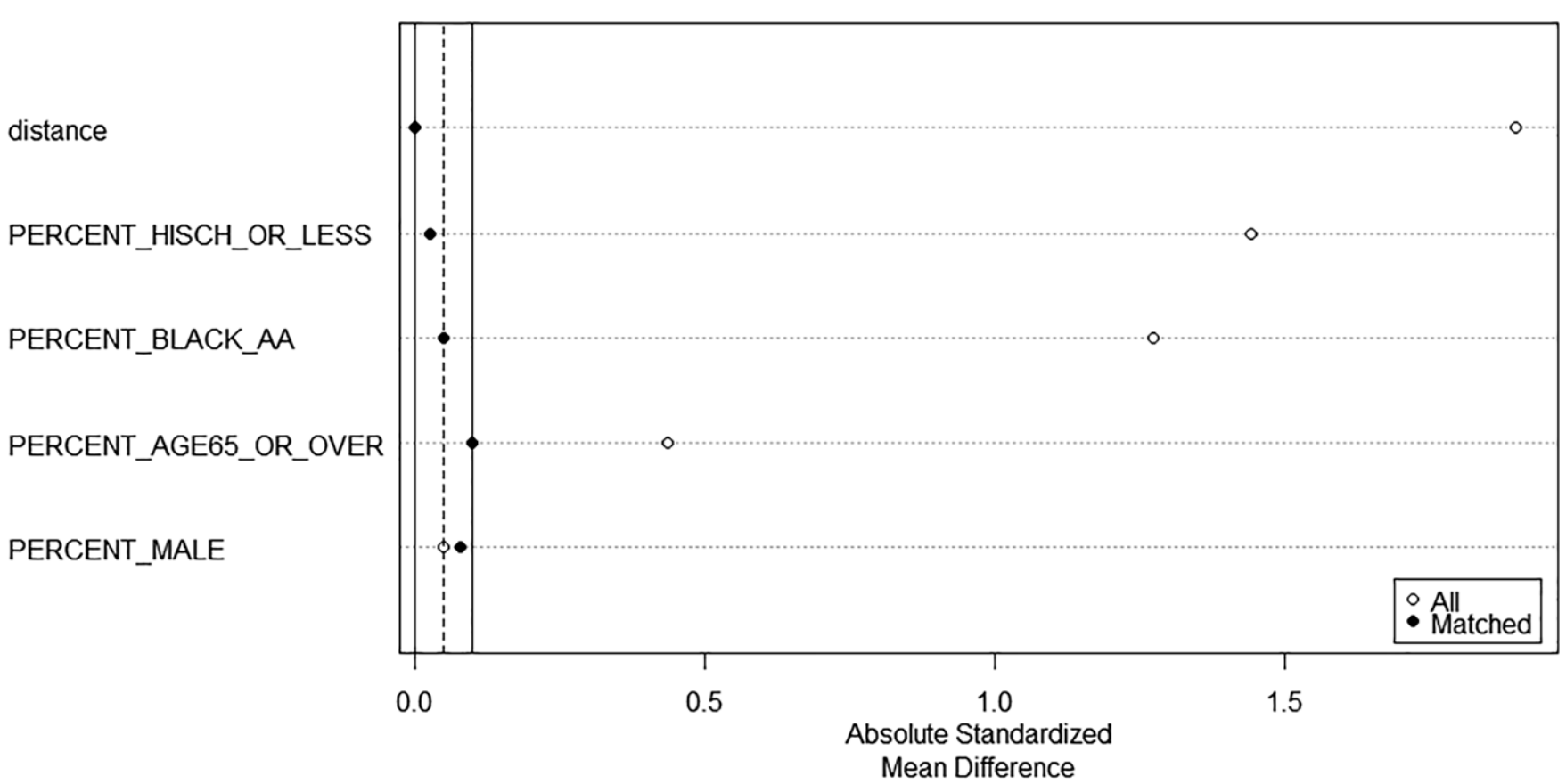

**FIGURE 7** Comparison of absolute standardized mean difference

effects are positive for both SDI and the staying-home percentage from the best performance and the effort perspectives and are negative for all other fourteen scenarios, which means that high income contributes to more desired social distancing performance in each of the 18 scenarios. As Table 4 reveals, ATE and ATET are very close in each scenario and both provide consistent conclusions, so only ATE is discussed below.

For example, ATE is 4.858 for the best performance of SDI, meaning on average being high income increases the highest SDI value by 4.858. ATE is 23.841% for the effort of the staying-home percentage, meaning on average being high income increases the percentage increase of staying-home percentage by 23.841%. ATE is −5.547% for the effort of miles traveled per person, meaning on average being high income decreases the percentage change (a negative value) of miles traveled per person by −5.547%.

From the effort perspective, a high income has the largest causal impact on the staying-home percentage. People from high-income communities were probably able to telework during the pandemic while this might not have

**TABLE 4** Causal effect of high income on the 18 scenarios

| Groups | Causal effect | Social distancing index | % staying home | Miles traveled/person | Trips/ person | Work trips/person | Non-work trips/person |
|---|---|---|---|---|---|---|---|
| Best performance | ATE | 4.858$^{***}$ | 2.255$^{***}$ | −1.259$^{***}$ | −0.188$^{***}$ | −0.028$^{***}$ | −0.128$^{***}$ |
| | ATET | 4.780$^{***}$ | 2.274$^{***}$ | −1.305$^{***}$ | −0.181$^{***}$ | −0.026$^{***}$ | −0.129$^{***}$ |
| Effort | ATE | 25.490$^{***}$ | 23.841$^{***}$ | −5.547$^{***}$ | −3.97$^{***}$ | −3.90$^{***}$ | −2.933$^{***}$ |
| | ATET | 22.809$^{***}$ | 22.046$^{***}$ | −5.586$^{***}$ | −3.66$^{***}$ | −3.85$^{***}$ | −2.748$^{***}$ |
| Consistency | ATE | −0.0610$^{***}$ | −0.330$^{***}$ | −0.593$^{***}$ | −0.034$^{***}$ | −0.014$^{***}$ | −0.031$^{***}$ |
| | ATET | −0.614$^{***}$ | −0.342$^{***}$ | −0.592$^{***}$ | −0.032$^{***}$ | −0.014$^{***}$ | −0.030$^{***}$ |

*Note*: significance level: 0.0001 - '***'.

been an option for people from low-income communities. Door-to-door services, such as food delivery, are more affordable for high-income people. A major impact of a high income on the staying-home percentage should encourage policymakers to analyze the underlying reasons and consider the living conditions of low-income communities during pandemics. Meanwhile, the expected impact of a high income on the percentage change (a negative value) of work-trip rate is greater than that of non-work trips. This indicates that the work trip rate is expected to further decrease compared with the non-work trip rate under the impact of high income. Teleworking is probably the major reason for a higher reduction in the work trip rate. This provides additional evidence regarding the social issue that low-income people were under higher risk of unemployment and were working in more exposed environments. Additionally, it is found that the causal impact of high income on the effort of miles traveled per person is larger than that of trip rate: −5.547% versus −3.97%. This reveals that high income also decreases the percentage change (a negative value) of the average travel distance per trip per person. Specifically, people in high-income communities are expected to conduct shorter trips. First, a higher accessibility to life supplies for high-income communities is probably a major reason for shorter non-work trips. Second, high-income people were more likely to have long-distance travel before the pandemic, while during the pandemic such long-distance travel (usually for work or tourism) was dramatically reduced, leading to shortened work or non-work trips.

It is expected that the best performance of miles traveled per person on average across all communities will be 1.259 miles less when all the communities are high income as opposed to when they are all low income. Although this causal impact is significant at the 99.9% confidence interval, its small value of 1.259 indicates that the improvement derived from high income is limited. In other words, the two income groups did not differ much in the minimum miles traveled per person.

Regarding consistency, high income reduces the variations of all six mobility metrics during the inertia and fatigue periods after the national emergency declaration on March 13, 2020. A more stable and consistent status is expected at the average level if the communities are high income than if all the communities are low income. For example, the ATE of high-income communities for the consistency of SDI is −0.610, which means that a high income is expected to reduce the SDI value's standard deviation by 0.610 on average.

Since the six mobility metrics do not have the same unit, some row-wise comparisons in Table 4, such as SDI and trip rate from a best performance perspective, are infeasible. Hence, the six metrics specifically in the best performance and effort aspects are first standardized and then undergo PSM analysis. The data regarding effort do not need standardization since all six metrics are in the same unit of percentage, and row-wise comparisons are discussed above. Table 5 summarizes the results after standardization. As far as best performance is concerned, high income has the largest impact on SDI, followed by trip rate, and then the non-work trip rate. The ATE on the non-work trip rate is much larger than on the work trip rate, which indicates that the non-work trip rate has more potential to be further reduced by high income. Regarding consistency, high income has a higher impact on the variation

**TABLE 5** Causal effect of high income after data standardization regarding best performance and consistency

| Groups | Causal effect | Social distancing index | % staying home | Miles traveled/ person | Trips/ person | Work trips/ person | Non-work trips/person |
|---|---|---|---|---|---|---|---|
| Best performance (extremum) | ATE | 0.473*** | 0.245*** | −0.191*** | −0.437*** | −0.198*** | −0.369*** |
| | ATET | 0.464*** | 0.245*** | −0.200*** | −0.427*** | −0.200*** | −0.350*** |
| Consistency (standard deviation) | ATE | −0.275*** | −0.177*** | −0.236*** | −0.310*** | −0.314*** | −0.293*** |
| | ATET | −0.283*** | −0.184*** | −0.222*** | −0.301*** | −0.291*** | −0.281*** |

*Note*: significance level: 0 '***' 0.001 '**' 0.01 '*' 0.05 '.' 0.1.

of the three trip-rate-related metrics: trip rate, work trip rate, and non-work trip rate. Trip-rate-related metrics are stabilized more easily by high income than the other three metrics—SDI, the staying-home percentage, and miles traveled per person.

## 4 | DISCUSSION

Mobility restrictions are undoubtedly crucial at the early stage of a pandemic. This is especially true of the fatigue period which could occur amid on-going stay-at-home orders; thus, early-stage mobility interventions should be as effective as possible. Although state and local governments issued various mobility interventions, such as stay-at-home orders, closing schools, and cancelling public events, the different effects of these interventions on communities of different income levels were rarely taken into consideration. This article investigates and expounds the disparities between income groups with respect to social distancing performances during the pandemic.

We identified real-world differences between communities in an unprecedented period of the COVID-19 pandemic. In comparison to an open-sourced database, we produced a much more granular database, specifically at census tract level across the country. Lots of details are shown by our multifaceted analysis from three aspects (best performance, effort, and consistency) and six mobility metrics (SDI, the staying-home percentage, miles traveled per person, trip rate, work trip rate, and non-work trip rate). Elaborated from all 18 scenarios, income plays a significant role in social distancing practices, and being high income promotes a community's social distancing performance, thus helping to contain the spread of the disease.

Many studies have shown that low-income neighborhoods report more COVID-19 cases and suffer higher death rates (Hong et al., 2021; Jay et al., 2020; Kikuchi et al., 2021; Truong & Asare, 2020), which put a heavier burden on containing the pandemic. One reason is that low-income people are more likely to have poor health, making them more vulnerable (Ghisolfi et al., 2020; Ivers & Walton, 2020; Whitehead et al., 2021). Our article provides an additional explanation as to why low-income communities have higher infection risks—because of their relatively higher travel needs. The possible reasons behind this are discussed. For example, high-income individuals are more likely to telework or work in a secure environment while low-income individuals probably cannot telework in the isolation of their homes so that they take more risks of getting infected. In addition, better accessibility to life supplies could also explain high-income communities' better social distancing performance, such as shorter distances traveled per person, a lower non-work trip rate, and higher staying-home percentage. Residents of high-income neighborhoods, many of which are in heavily commercialized areas, can acquire basic necessities without traveling as far as low-income people. Additionally, home delivery services are increasingly popular, although this market is split between high- and low-income individuals. As found in the inertia and fatigue periods, low-income communities' social distancing performance fluctuated more than high-income communities'. Low-income people presumably had to leave their homes in order to make a living, even though partial societal reopening had not yet begun. This is supported in our research, which demonstrates that from a consistency standpoint, having a high salary has the biggest effect on the three trip-rate-related metrics (trip rate, work trip rate, and non-work trip rate). High income has more potential to stabilize daily travel frequency than the staying-home percentage. Analyzing the travel purposes of these more frequent trips may assist policymakers in better containing pandemics. Weill et al. (2020) analyzed several non-work travel purposes during the pandemic and found that supermarkets, carryout restaurants, and convenience stores were the most-visited destinations in April 2020. Their research also confirms our finding that low-income communities conduct relatively longer non-work trips after the pandemic emerged due to lower accessibility to daily supplies.

There are also some other studies investigating the disparity between income groups in mobility patterns during the pandemic. Coven and Gupta (2020) analyzed the role of income in the mobility difference in New York City. They also found that low-income people presented higher work activity after the pandemic. Weill et al. (2020) designed four metrics that differ from ours to discuss the difference across five quintile income groups. Their core finding was that low-income communities showed lower levels of social distancing. This is also our major finding, but our article established this conclusion using different metrics and from three different aspects. Iio et al. (2021) illustrated the disparities in mobility flexibility between income groups during the pandemic through a case study of Greater Houston, Texas. Much more detailed mobility metrics were designed, which are a good reference for small-scale studies but are too computationally costly to be applied to a nationwide analysis. Their finding of lower mobility flexibility among low-income people is consistent with our analysis that low-income people have reduced accessibility to daily necessities.

In summary, this article uses real-world observations to demonstrate how communities with different income levels responded to social distancing in a variety of ways. The rationale for each finding prompts discussion and raises concerns about the actual situation of low-income communities. There is room for improving this article in the future. Given the high rate of child poverty, the percentage of adults aged 65 years and older is chosen as the variable for logistic regression analysis rather than the percentage of children, which would be a preferable option. Other matching techniques besides nearest neighbor matching with a replacement might result in a more balanced outcome. Furthermore, assessing crowd concentration at certain locations, such as supermarkets and gyms, may be a more direct method of measuring social distancing performance than the six mobility indicators. Hotspot analysis, which monitors crowd concentration levels and identifies densely populated areas, may be used in the future to assist policymakers in implementing the proper steps during the early stages of a pandemic. Investigating the differences in personal behavior between communities based on income level also merits future study.

## ORCID

*Qianqian Sun* https://orcid.org/0000-0003-3684-4603
*Aliakbar Kabiri* https://orcid.org/0000-0003-2119-007X